\newif
\ifconfver
\confvertrue        
\ifconfver
    \documentclass[10pt,journal,final,twoside]{IEEEtran}
\else
    \documentclass[11pt,draftcls,onecolumn,draftclsnofoot]{IEEEtran}
    \usepackage{setspace}
\fi
\usepackage{xspace,empheq,fancybox,amsmath,amssymb,graphicx,epstopdf,epsfig,syntonly,times,amsthm} \usepackage{psfrag,color,bm,array,cite}
\usepackage{url,cite,footnote,xspace,syntonly,algorithm,algorithmic}
\usepackage{verbatim,multirow}
\usepackage{tabularx}
\usepackage[T1]{fontenc}
\usepackage{mathtools}
\usepackage{amsmath,amsfonts,amsthm,bm} 
\usepackage{graphicx}
\usepackage{mwe}
\usepackage{caption}
\usepackage{cite}
\usepackage{amsmath,amssymb,amsfonts}
\usepackage{algorithmic}
\usepackage{graphicx}
\usepackage{textcomp}
\usepackage{xcolor}
\usepackage{multicol, blindtext}
\usepackage{xspace,empheq,fancybox,amsmath,amssymb,graphicx,epstopdf,epsfig,syntonly,times,amsthm} \usepackage{psfrag,color,bm,array,cite}
\usepackage{url,cite,footnote,xspace,syntonly,algorithm,algorithmic}
\usepackage{verbatim,multirow}
\usepackage[T1]{fontenc}
\usepackage{mathtools}
\usepackage{amsmath,amsfonts,amsthm,bm}
\usepackage{graphicx}
\usepackage{mwe}
\usepackage{caption}
\usepackage{stfloats}
\usepackage{amsfonts}
\usepackage{stackengine}
\usepackage{lipsum}
\usepackage{bbm}

\usepackage{xcolor}
\usepackage{array}
\usepackage{seqsplit}
\makeatletter
\newcommand{\removelatexerror}{\let\@latex@error\@gobble}
\makeatother

\newcolumntype{P}[1]{>{\centering\arraybackslash}p{#1}}

\usepackage{nomencl}

\usepackage{diagbox}
\usepackage{etoolbox}
\usepackage{enumitem}

\makenomenclature

\usepackage{hhline}
\newtheorem{remark}{\bfseries Remark}
\input{mysymbol.sty}

\definecolor{orange}{rgb}{1,0.5,0}

\definecolor{color_yuqi}{RGB}{0, 0, 255}

\definecolor{color_yuqi2}{RGB}{234, 135, 47}

\allowdisplaybreaks

\makeatletter
\let\old@ps@headings\ps@headings
\let\old@ps@IEEEtitlepagestyle\ps@IEEEtitlepagestyle
\def\psccfooter#1{%
    \def\ps@headings{%
        \old@ps@headings%
        \def\@oddfoot{\strut\hfill#1\hfill\strut}%
        \def\@evenfoot{\strut\hfill#1\hfill\strut}%
    }%
    \def\ps@IEEEtitlepagestyle{%
        \old@ps@IEEEtitlepagestyle%
        \def\@oddfoot{\strut\hfill#1\hfill\strut}%
        \def\@evenfoot{\strut\hfill#1\hfill\strut}%
    }%
    \ps@headings%
}
\makeatother

\usepackage{nomencl}

\usepackage{ifthen}

\usepackage{subcaption}
\begin{document}

\title{Equitable Networked Microgrid Topology Reconfiguration for Wildfire Risk Mitigation}
\author{
\IEEEauthorblockN{Yuqi Zhou},~\IEEEmembership{Member, IEEE}, \IEEEauthorblockN{Ahmed S. Zamzam},~\IEEEmembership{Member, IEEE}, and \IEEEauthorblockN{Andrey Bernstein},~\IEEEmembership{Member, IEEE}

\thanks{\protect\rule{0pt}{0mm} 
This work was authored by the National Renewable Energy Laboratory, operated by Alliance for Sustainable Energy, LLC, for the U.S. Department of Energy (DOE).
The views expressed in the article do not necessarily represent the views of the DOE or the U.S. Government. The U.S. Government retains and the publisher, by accepting the article for publication, acknowledges that the U.S. Government retains a nonexclusive, paid-up, irrevocable, worldwide license to publish or reproduce the published form of this work, or allow others to do so, for U.S. Government purposes.
}

}



\renewcommand{\thepage}{}
\maketitle
\pagenumbering{arabic}

%

\begin{abstract}

The increasing number of wildfires in recent years consistently challenges the safe and reliable operations of power systems. To prevent power lines and other electrical components from causing wildfires under extreme conditions, electric utilities often deploy public safety power shutoffs (PSPS) to mitigate the wildfire risks therein. Although PSPS are effective countermeasures against wildfires, uncoordinated strategies can cause disruptions in electricity supply and even lead to cascading failures. Meanwhile, it is important to consider mitigating biased decisions on different communities and populations during the implementation of shutoff actions. In this work, we primarily focus on the dynamic reconfiguration problem of networked microgrids with distributed energy resources. In particular, we formulate a rolling horizon optimization problem allowing for flexible network reconfiguration at each time interval to mitigate wildfire risks. To promote equity and fairness during the span of shutoffs, we further enforce a range of constraints associated with load shedding to discourage disproportionate impact on individual load blocks. Numerical studies on a modified IEEE 13-bus system and a larger-sized Smart-DS system demonstrate the performance of the proposed algorithm towards more equitable power shutoff operations.

\end{abstract}




\section{Introduction}\label{sec:intro}

{R}{ecent} years have witnessed a surge of interest in the development of microgrids, thanks to the increasing renewable penetration and advances in energy storage techniques. With the evolution of remote communication and control, networked microgrids become possible where multiple individual microgrids are allowed to be connected to each other. The interconnectivity within the networked system helps facilitate resource sharing and also improves grid resilience against natural disasters such as wildfires. To mitigate the wildfire risks, public safety power shutoffs (PSPS) are commonly implemented by utility companies, to reduce the possibility of electrical components (e.g., power lines, transformers, insulators) causing fires due to misoperations or extreme weather conditions.
Admittedly, proactively shutting off power reduces the risk of wildfire ignitions and improves system resilience in the long run. However, power outages would definitely cause inconvenience for customers, and even raise safety and health concerns. Experiencing prolonged and frequent load shedding is never desired for end users \cite{lacommare2004understanding}.
To maximize load delivery, electrical components such as power lines can be switched strategically, which allows for more flexible network topology and enhanced transfer capabilities. 
In fact, wildfires tend to happen more frequently within distribution networks (including microgrids) than in transmission systems \cite{xcel}, and the cost of installing controllable switches in microgrids is much lower than in transmission systems \cite{backhaus2015dc,chen2020networked}.
This motivates the design of rolling-horizon topology reconfiguration strategies specifically for networked microgrids that also account for equity and fairness concerns.

Earlier work on PSPS mainly focuses on obtaining the optimal network topology and load shedding plans. Optimization problems are formulated primarily to maximize load delivery \cite{coffrin2018relaxations,zhou2023optimal,lesage2023optimally}, minimize wildfire risks \cite{rhodes2020balancing,astudillo2022managing,taylor2022framework,su2023quasi,yang2024multi}, or improve grid resilience \cite{ma2016resilience,liu2016microgrids,trakas2017optimal,abdelmalak2022enhancing}. However, the equity and fairness concerns have been overlooked in these previous works.
In fact, the research on PSPS with equity and fairness considerations is fairly limited. 
The optimal design of networked microgrids to manage wildfire risks with equity considerations is studied in \cite{taylor2023managing}. The vulnerability of customers has been considered in the objective function, to discourage biased load shedding. Nonetheless, the problem is formulated as a single-step optimization, which fails to consider the long-term equity of end customers. 
To address this, \cite{kody2022sharing} presents a rolling horizon optimization problem for mitigating wildfire risk while promoting fairness of load shedding. However, the primary focus of this work centers on the operation of transmission systems rather than directly addressing the fairness of residential customers. In addition, the formulations lack comparisons among load shedding across different buses to ensure equity.
More recently, \cite{sundar2023fairly} has put forth approaches to incorporate fairness into the minimum load shedding problem, but it is primarily designed for transmission systems and has not addressed fairness over long-term operational periods. Therefore, developing an equitable rolling-horizon PSPS algorithm on networked microgrid systems is in pressing need, and the research domain is still open for exploration.

The goal of this paper is to design a rolling horizon topology reconfiguration algorithm for networked microgrids that can effectively mitigate wildfire risk while accounting for the equity of the load shedding decisions.
To achieve this, we approach by first formulating a rolling horizon optimal network reconfiguration problem, for which the load blocks and switches are allowed to open/close flexibly to minimize the total load shedding. With the support from grid-forming (-following) inverters and distributed energy resources, the microgrids can either connect to the main grid or operate on their own in island mode.
To attain more equitable topology reconfiguration decisions, we further enforce diverse constraints on load shedding over load blocks. Specifically, these constraints play important roles in identifying emergency loads, hedging against prolonged/frequent power disruptions, and preventing disproportionate shutoffs.
Our results show that the equitable reconfiguration formulation provides an improved strategy, which can effectively mitigate the biased impact of load shedding on any particular group of customers.

This paper is organized as follows. Section \ref{sec:original} starts with the modeling of distribution systems, and presents the optimal network reconfiguration problem for wildfire mitigation. Section \ref{sec:equity} presents the equitable reconfiguration formulation, by further accounting for diverse equity constraints on load shedding operations. Numerical simulations on a modified IEEE 13-bus system and a larger SmartDS system are presented in Section \ref{sec:num} to demonstrate the performance of the algorithm and the paper is concluded in Section \ref{sec:con}.


\makenomenclature

\renewcommand\nomgroup[1]{%
  \item[\bfseries
  \ifstrequal{#1}{S}{Sets}{%
  \ifstrequal{#1}{C}{Constants}{%
  \ifstrequal{#1}{D}{Decision Variables (Optimization)}{}}}%
]}

\nomenclature[S]{\({\cal L}\)}{Set of lines}
\nomenclature[S]{\({{\cal L}_{\text{sw}}}\)}{Set of switchable lines}
\nomenclature[S]{\({\cal N}\)}{Set of buses}
\nomenclature[S]{\({\cal N}_{i}\)}{Neighboring nodes of $i$}

\nomenclature[S]{\({\cal T}\)}{Set of time intervals}

\nomenclature[s]{\(\Phi\)}{Set of phases}

\nomenclature[s]{\(\mathcal B\)}{Set of load blocks}

\nomenclature[s]{\(\cal M\)}{Set of emergency blocks}

\nomenclature[s]{\(\mathcal{B}\setminus\mathcal{M}\)}{Set of non-emergency blocks}

\nomenclature[s]{\({\cal G}_{i}\)}{Set of generators connected to node $i$}

\nomenclature[s]{\({\cal D}_{i}\)}{Set of loads connected to node $i$}

\nomenclature[s]{\({\cal G}_{i}^{s}\)}{Set of energy storage units connected to node $i$}

\nomenclature[s]{\({\cal W}_{k}\)}{Set of DERs within connected component $k$}

\nomenclature[s]{\({\cal S}_{k}\)}{Set of switches linked to connected component $k$}

\nomenclature[D]{\(z_{\kappa}\)}{Energization of load blocks}

\nomenclature[D]{\(z^{\text{inv}}_{i}\)}{Inverter object status (grid forming or following)}

\nomenclature[D]{\(w_{i}\)}{Squared voltage magnitudes}

\nomenclature[D]{\(S^{g}_{i}\)}{Complex generation power}

\nomenclature[D]{\(S^{d}_{i}\)}{Complex load power}

\nomenclature[D]{\(S_{i}\)}{Complex net power injection}

\nomenclature[D]{\(P^{g}_{i}, Q^{g}_{i}\)}{Real/reactive generation power}

\nomenclature[D]{\(P^{d}_{i}, Q^{d}_{i}\)}{Real/reactive load power}

\nomenclature[D]{\(P_{i}, Q_{i}\)}{Real/reactive net power injections}

\nomenclature[D]{\(E_{i}\)}{Energy level of storage unit}

\nomenclature[D]{\(z^{s}_{i}\)}{Energy storage operational status}

\nomenclature[D]{\({P}^{+}_{i}, {P}^{-}_{i}\)}{Charging/discharging power}

\nomenclature[D]{\(P_{ij}, Q_{ij}\)}{Real/reactive line flow power}

\nomenclature[D]{\(S_{ij}\)}{Complex line flow power}

\nomenclature[D]{\(z^{+}_{i}, z^{-}_{i}\)}{Charging/discharging status of storage unit}

\nomenclature[D]{\(\Delta z_{\kappa}\)}{Change of load block status}

\nomenclature[D]{\(z^{\text{sw}}_{\ell}\)}{Status of switchable lines}



\nomenclature[D]{\(\varphi_{ij}\)}{Directional binary variable defined on path $(i,j)$}

\nomenclature[D]{\(\zeta_{ij}\)}{Auxiliary binary variable on the status of $(i,j)$}

\nomenclature[D]{\(f^{k}_{ij}\)}{Fictitious commodity flowing on path $(i,j)$}

\nomenclature[D]{\({L}_{ij}\)}{Term of total power loss on $(i,j)$}

\nomenclature[C]{\(\eta^{+}_{i}, \eta^{-}_{i}\)}{Charging/discharging efficiency}

\nomenclature[C]{\(r_{ij}, x_{ij}\)}{Line resistance/reactance}





\nomenclature[C]{\(r_{\kappa}\)}{Load block wildfire ignition risk}

\nomenclature[C]{\(\epsilon\)}{Wildfire risk safety threshold}

\nomenclature[C]{\(m\)}{Limit on frequency of load shedding}

\nomenclature[C]{\(\lambda\)}{Maximum ratio of load shedding}

\nomenclature[C]{\({\alpha}_{\kappa}\)}{Limit on total number of load shedding}

\nomenclature[C]{\(v_{\kappa}\)}{Vulnerability index of load block}

\nomenclature[C]{\(\psi_{\kappa}\)}{Percentage threshold on load shedding}

\nomenclature[C]{\(\beta_{\kappa\nu}\)}{Limit on the ratio of load shedding}

\nomenclature[C]{\(\rho\)}{Weighting factor of dynamic reconfiguration problem}

\nomenclature[C]{\(R^{g}_{u}, R^{g}_{d}\)}{Ramp up and down limits for generation units}

\nomenclature[C]{\(\overline{k}_{\text{bl}}\)}{Switching limits for load blocks}

\nomenclature[C]{\(\overline{k}_{\text{sw}}\)}{Switching limits for switchable lines}

\printnomenclature 

\vspace{2mm}













\section{Optimal Network Reconfiguration Problem}
\label{sec:original}


We start with modeling the networked microgrid. Let us consider a steady state distribution system, with $N$ buses collected in the set $\cal N :=$ $\{ 1,\ldots,N \}$, and $L$ branches defined in the set $\cal L :=$ $\{(i,j)\}$. 
In addition, we use $\mathcal{L}_{\text{sw}} \subseteq \mathcal{L}$ to denote the set of switchable lines.
The complex power of generation and load connected to node $i$ is  denoted by $S^{g}_{i} = P^{g}_{i} + j Q^{g}_{i}$ and $S^{d}_{i} = P^{d}_{i} + j Q^{d}_{i}$, respectively.
During network reconfiguration or emergency operations, due to limited network transfer capability, certain loads need to be shed to maintain system power balance. Different from network reconfiguration in transmission systems, in distribution systems, the loads are usually served or shed simultaneously across all nodes within a portion of the network.
The term ``load block'' is used to describe a set of connected components in the network that remain active when all switchable lines are disconnected. Hence, the load block status can be defined using a binary decision variable $z_{\kappa,t}$, for which $z_{\kappa,t} = 1$ means a load block $\kappa$ is energized at time $t$ and $z_{\kappa,t} = 0$ represents de-energization of a load block.
Our main focus in this work is to address a network operational planning problem, to eventually determine the connection of load blocks across the rolling horizon indicated by $\cal T :=$ $\{ 1, 2, \ldots, T \}$. To this end, we will next introduce the related constraints and the optimization formulation for the network reconfiguration problem.

\vspace{-0.2cm}

\subsection{Power Flow Constraints}
The power flow of the distribution system is formulated in a generalized form \cite{baran1989optimal,li2012exact} as follows. For any line $(j,k) \in \cal L$ at time $t$, the following equation holds:
\begin{align}
    {w}_{k,t} = {w}_{j,t} - 2\left({r_{jk}}P_{jk,t} + {x_{jk}}Q_{jk,t}\right) + {{L}_{jk,t}},  \: \forall t \in \mathcal T  \label{eq:pf_1}
\end{align}
in which ${w}_{t}$ denotes the squared voltage magnitude at the corresponding node. Additionally, ${{L}_{jk,t}}$ is related to losses and equals $\left(r^{2}_{jk} + x^{2}_{jk}\right) \mathit{l}_{jk}$, where $\mathit{l}_{jk} = \frac{P^{2}_{jk,t} + Q^{2}_{jk,t}}{w_{j,t}}$.
Per nodal power balance, the total line flow into node $j$ is given by:
\begin{align}
    {S}_{ij,t} =  \sum_{(j,k) \in \cal L}S_{jk,t} - {S}_{j,t},  \quad \forall t \in \mathcal T  \label{eq:pf_2}
\end{align}
where ${S}_{j,t} = {S}^{g}_{j,t} - {S}^{d}_{j,t}$ denotes the complex net power injection at node $j$.
\begin{remark}[Power flow modeling]
The power flow equations for distribution systems are presented in a generalized form above. Various approximation methods (e.g., linear \cite{arnold2016optimal,bernstein2017linear}, semidefinite \cite{lavaei2013geometry,huang2022three} models) can be leveraged to streamline these formulations, which enables more tractable computation.
\end{remark}

\vspace{-0.4cm}


\subsection{System Operation Constraints}

Constraints on voltage magnitude for each node $i$ at time step $t$ are enforced by:
\begin{align}
\underline{V}^{2}_{i,t}z_{\kappa,t} \leq w_{i,t} \leq \overline{V}^{2}_{i,t}z_{\kappa,t},  \quad \forall t \in \mathcal T \label{eq:con_voltage}
\end{align}
where $\kappa$ is the load block that $i$ belongs to, and $\underline{V}_{i,t}, \overline{V}_{i,t}$ are voltage limits. When the block is energized, the constraint enforces lower and upper limits on the non-negative voltage magnitude. On the other hand, for the de-energized block, it boils down to a single equality constraint $w_{i,t} = 0$.

Given the real generation limits $\underline{P}^{g}_{i,t}, \overline{P}^{g}_{i,t}$ and reactive generation limits $\underline{Q}^{g}_{i,t}, \overline{Q}^{g}_{i,t}$, the constraints on generation for each node $i$ at time step $t$ can be formulated as:
\begin{subequations} \label{eq:con_gen}
\begin{align}
    & \underline{P}^{g}_{i,t} z_{\kappa,t} \leq {P}^{g}_{i,t} \leq \overline{P}^{g}_{i,t} z_{\kappa,t},  \quad \forall t \in \mathcal T\\
    & \underline{Q}^{g}_{i,t} z_{\kappa,t} \leq Q^{g}_{i,t} \leq \overline{Q}^{g}_{i,t} z_{\kappa,t},  \quad \forall t \in \mathcal T
\end{align}
\end{subequations}
Similarly, constraints on loads at node $i$ are given by:
\begin{subequations} \label{eq:con_load}
\begin{align}
    & \underline{P}^{d}_{i,t} z_{\kappa,t} \leq P^{d}_{i,t} \leq \overline{P}^{d}_{i,t} z_{\kappa,t},  \quad \forall t \in \mathcal T\\
    & \underline{Q}^{d}_{i,t} z_{\kappa,t} \leq Q^{d}_{i,t} \leq \overline{Q}^{d}_{i,t} z_{\kappa,t},  \quad \forall t \in \mathcal T
\end{align}
\end{subequations}
The nonzero upper/lower bounds for generation and load are only applied when the associated load block is energized; otherwise, the variables are restricted to zero. Non-dispatchable generators and loads can be also conveniently modeled by setting identical upper and lower bounds. In addition, the ramping constraints for generation are given by:
\begin{subequations} \label{eq:con_ramping}
\begin{align}
    & P^{g}_{i,t} - P^{g}_{i,t-1} \leq R^{g}_{u},  \quad \forall t = 2, \ldots, T\\
    & P^{g}_{i,t-1} - P^{g}_{i,t} \leq R^{g}_{d},  \quad \forall t = 2, \ldots, T
\end{align}
\end{subequations}
where $R^{g}_{u}, R^{g}_{d}$ denote ramping up and down limits, and these constraints can regulate the rate at which generation increases or decreases between time periods.

For flexible network reconfiguration, a subset of lines ${\cal L}_{\text{sw}} \subseteq \cal L$ can be switched on/off at each time interval. The status of each individual line $\ell = (i, j) \in \cal L$ can be represented using a binary variable $z_{\ell, t}$. For switchable lines $\ell \in {\cal L}_{\text{sw}}$, $z_{\ell, t}$ can be either 1 (connected) or 0 (disconnected), whereas for non-switchable lines $\ell \in \cal{L}\setminus{\cal L}_{\text{sw}}$ the line status $z_{\ell, t}$ is restricted to 1 during the rolling horizon:
\begin{subequations} \label{eq:con_switchable}
\begin{align}
    &  0 \leq z_{\ell,t} \leq  1,  \quad \forall \ell \in {\cal L}_{\text{sw}}, \quad \forall t \in \mathcal T\\
    & {z}_{\ell,t} = 1,  \quad \forall \ell \in {\cal{L}}\setminus{\cal L}_{\text{sw}}, \quad \forall {t} \in \mathcal T
\end{align}
\end{subequations}
To this end, the constraints on real and reactive line flows are enforced by:
\begin{subequations} \label{eq:con_branch}
\begin{align}
    &  \underline{P}_{ij,t} z_{ij,t} \leq P_{ij,t} \leq  \overline{P}_{ij,t} z_{ij,t},  \quad \forall t \in \mathcal T\\
    & \underline{Q}_{ij,t} z_{ij,t} \leq Q_{ij,t} \leq \overline{Q}_{ij,t} z_{ij,t},  \quad \forall t \in \mathcal T
\end{align}
\end{subequations}
Furthermore, nodal power balance leads to the following:
\begin{align}
     \sum_{k \in {\cal G}_{i}}S^{g}_{k,t} - \sum_{k \in {\cal D}_{i}}  S^{d}_{k,t} = \sum_{j \in {\cal N}_{i}} S_{ij,t},  \quad \forall t \in \mathcal T
     \label{eq:con_balance}
\end{align}
where ${\cal G}_{i}$ and ${\cal D}_{i}$ denote the set of generators and loads connected to node $i$, and we use ${\cal N}_{i}$ to denote the set of neighboring nodes of $i$. 

\vspace{-0.5cm}

\subsection{Energy Storage Constraints}
The networked microgrid system is also equipped with energy storage devices, to balance load and support grid stability. For any energy storage unit $i \in {\cal G}_{i}^{s}$, its energy levels yield the following constraints:
\begin{subequations} \label{eq:eq_storage}
\begin{align}
    & E_{i,t} - E_{i,t-1} = \eta^{+}_{i} P^{+}_{i,t} - \frac{P^{-}_{i,t}}{\eta^{-}_{i}},  \; \forall t = 2, \ldots, T\\
    & 0 \leq E_{i,t} \leq \overline{E}_{i},  \quad \forall t \in \mathcal T
\end{align}
\end{subequations}
where $P^{+}_{i,t}, P^{-}_{i,t}$ denote the charging and discharging power at time interval $t$, and constants $\eta^{+}_{i}, \eta^{-}_{i}$ denote the charging and discharging efficiency, respectively. To this end, we further introduce binary variables $z^{+}_{i,t}, z^{-}_{i,t}$ for energy storage unit charging/discharging status, leading to the following complementary constraints:
\begin{subequations} \label{eq:eq_storage_complementary}
\begin{align}
    & z^{+}_{i,t} + z^{-}_{i,t} = z^{s}_{i,t},  \quad \forall t \in \mathcal T \label{eq:eq_storage_complementary_a}\\
    & 0 \leq P^{+}_{i,t} \leq z^{+}_{i,t} \overline{P}^{+}_{i},  \quad \forall t \in \mathcal T \label{eq:eq_storage_complementary_b}\\
    & 0 \leq P^{-}_{i,t} \leq z^{-}_{i,t} \overline{P}^{-}_{i},  \quad \forall t \in \mathcal T \label{eq:eq_storage_complementary_c}
\end{align}
\end{subequations}
The energy storage operational status is represented by a $0/1$ variable $z^{s}_{i,t}$ in \eqref{eq:eq_storage_complementary_a}. Further, the constraints on charging and discharging power are enforced in \eqref{eq:eq_storage_complementary_b} and \eqref{eq:eq_storage_complementary_c}, where $\overline{P}^{+}_{i}, \overline{P}^{-}_{i}$ represent the charging/discharging power ratings.

\vspace{-0.5cm}

\subsection{Wildfire Mitigation Constraints}
The network reconfiguration problem can be effectively employed to model an optimization framework in order to mitigate wildfire risks. This can be achieved by limiting the total wildfire risk to be below a specified safety threshold. Given the wildfire ignition risk $r_{\kappa,t}$ of each load block $\kappa$ at time interval $t$, the constraint can be accordingly cast as:
\begin{align}
    {\sum_{\kappa \in \mathcal B}r_{\kappa,t}z_{\kappa,t}} \leq \epsilon \sum_{\kappa \in \mathcal B}r_{\kappa,t}, \quad \forall t \in \mathcal T  \label{eq:con_fire}
\end{align}
where $\mathcal B$ denotes the set of load blocks and $\epsilon \in [0,1]$ is the predetermined safety threshold. The constraint ensures that the total wildfire risks for energized load blocks cannot exceed the maximum during any time interval $t$. Of course, the above formulation can be easily modified to address a wide range of risk mitigation problems that arise in power systems.
\begin{remark}[Wildfire risk modeling]
The wildfire risk $r_{\kappa,t}$ can be obtained using Wildland Fire Potential Index (see e.g., \cite{WFPI,vazquez2022wildfire,yang2024multi}), which provides the relative wildfire potential due to various factors such as wind, moisture, and vegetation.
\end{remark}

\subsection{Network Topology Constraints}
Extensive network switching can cause disturbances to the system and raise stability concerns, therefore operational limits on the total number of blocks and lines switched are enforced, as given by:
\begin{align}
    & \sum_{\kappa \in \mathcal B} (1-z_{\kappa,t}) \leq \overline{k}_{\text{bl}},  \quad \forall t \in \mathcal T \label{eq:switch_limit_block}\\
    & \sum_{\ell\in\mathcal{L}_{\text{sw}}} (1-z^{\text{sw}}_{\ell,t}) \leq \overline{k}_{\text{sw}},  \quad \forall t \in \mathcal T. \label{eq:eq_switch_limit_line}
\end{align}
Here, $\overline{k}_{\text{bl}}$ and $\overline{k}_{\text{sw}}$ represent the switching limits for load blocks and switchable lines, respectively. Notably, adding these switching budgets can also reduce the search space and significantly improve the computational efficiency (e.g., \cite{hedman2008optimal,zhou2022distributionally}) of the resulting optimization problem.

To preserve the integrity of the network structure during each time interval, multiple topology constraints need to be enforced. We start by imposing a topology constraint to ensure the statuses of any pair of neighboring blocks ($\kappa_i$ and $\kappa_j$) are aligned with the switch status ($z^{\text{sw}}_{\ell}$) between them:
\begin{subequations} \label{eq:con_topology1}
\begin{align}
    &  z_{\kappa_i,t} - z_{\kappa_j,t} \leq 1-z^{\text{sw}}_{\ell,t}, \: \forall \ell = (\kappa_i,\kappa_j), \: \forall t \in \mathcal T\\
    &  z_{\kappa_i,t} - z_{\kappa_j,t} \geq z^{\text{sw}}_{\ell,t} - 1, \: \forall \ell = (\kappa_i,\kappa_j), \: \forall t \in \mathcal T.
\end{align}
\end{subequations}
Furthermore, to ensure the system after reconfiguration follows a radial topology, we introduce the following constraints for the formation of spanning trees:
\begin{align} \label{eq:con_topology2}
    &  \sum_{(i,j) \in {\cal{L}} } (\varphi_{ij,t} + \varphi_{ji,t}) = |{\cal{N}}| - 1, \: \forall t \in \mathcal T\\
    & \varphi_{ij,t} + \varphi_{ji,t} = \zeta_{ij,t},  \: \forall t \in \mathcal T\\
    &  z^{\text{sw}}_{ij,t} \leq \zeta_{ij,t},  \: \forall t \in \mathcal T
\end{align}
in which $\varphi_{ij}$ is a directional binary variable defined on path $(i,j)$, and $\zeta_{ij}$ is an auxiliary binary variable representing the connection status of branch $(i,j)$. In addition, for all the nodes $k$ in set $\cal N$ that are not substation nodes (e.g., $k \in {\cal{N}} \backslash i_r$), the following directed multi-commodity flow \cite{lei2020radiality} based constraints are enforced:
\begin{align} \label{eq:con_topology3}
    &  \sum_{(j,i_r) \in {\cal{L}}} f^{k}_{j i_r,t} - \sum_{(i_r,j) \in {\cal{L}}} f^{k}_{i_r j,t} = -1, \: \forall t \in \mathcal T\\
    &  \sum_{(j,k) \in {\cal{L}}} f^{k}_{j k,t} - \sum_{(k,j) \in {\cal{L}}} f^{k}_{k j,t} = 1, \: \forall t \in \mathcal T\\
    &  \sum_{(j,i) \in {\cal{L}}} f^{k}_{j i,t} - \sum_{(i,j) \in {\cal{L}}} f^{k}_{i j,t} = 0, \: \forall t \in \mathcal T\\
    &  0 \leq f^{k}_{ij, t} \leq \varphi_{ij,t}, \ 0 \leq f^{k}_{ji, t} \leq \varphi_{ji,t}, \: \forall t \in \mathcal T.
\end{align}
Essentially, these constraints enforce $1$ fictitious unit of commodity to flow from the substation node to node $k$. Meanwhile, each commodity is only allowed to flow on a path if it is included in the directed spanning tree, as defined earlier.

\subsection{Inverter Constraints}
For a networked microgrid, we also need to make sure that after reconfiguration, every connected component is equipped with one grid-forming distributed energy resource (DER). Let ${\cal{W}}_{k}$ denote the set of DERs within connected component $k$, and ${\cal{S}}_{k}$ denote the set of switches connected to $k$. Accordingly, the following constraints are presented:
\begin{align} \label{eq:con_topology4}
    1 - \sum_{ij \in {\cal{S}}_{k}} z^{\text{sw}}_{ij,t} \leq \sum_{i \in {\cal{W}}_{k}} z^{\text{inv}}_{i,t}  \leq 1, \: \forall t \in \mathcal T
\end{align}
which indicates that there can be no more than one grid-forming DER per component, and an energized component without grid-forming inverters is allowed only if it is connected to components containing grid-forming inverters. In addition, the following constraint is enforced for each DER $i$ in the connected component:
\begin{align} \label{eq:con_topology5}
   S^{g}_{i,t} \leq \overline{S}^{g}_{i,t} \left(\sum_{ij \in {\cal{S}}_{k}} z^{\text{sw}}_{ij,t} + \sum_{i \in {\cal{W}}_{k}} z^{\text{inv}}_{i,t} \right), \: \forall t \in \mathcal T
\end{align}
which states that if all incident switches are open ($z^{\text{sw}}_{ij,t} = 0$), and the DER is grid-following ($z^{\text{inv}}_{i,t} = 0$), then the generation output are restricted to be zero. Otherwise, it does not affect the original constraint and becomes redundant. To transform nodal constraints into load block-based constraints might necessitate additional coloring constraints, and interested readers can refer to  PowerModelsONM \cite{fobes2022} for more details.

\begin{itemize}[label={$\bullet$}]
    \item \textit{Optimal Network Reconfiguration Problem} 
\end{itemize}

Given the constraints introduced above, the problem of optimal network reconfiguration can be formulated as:
\begin{subequations} \label{eq:original}
\begin{align}
\min \quad & \sum_{t \in \mathcal T} \sum_{\kappa \in \mathcal B} {P^{d}_{\kappa,t} (1-z_{\kappa,t})} \label{eq:O_a}\\
\textrm{s.t.} \quad &  \eqref{eq:pf_1} - \eqref{eq:con_topology5}   \label{eq:O_b}
\end{align}
\end{subequations}
which seeks to find the optimal network topology for each time interval $t$, while minimizing the total load shedding. The optimization problem can be conveniently implemented using the PowerModelsONM \cite{9897093} package in Julia.

\begin{remark}[Uncertainty of parameters]
Note that our formulations so far assume a deterministic model, but some parameters such as loads may be uncertain during real-time operations. Multiple mathematical models can be leveraged to handle the uncertainty in the rolling horizon formulation. These include chance-constrained programming \cite{qiu2014chance}, dynamic programming \cite{bellman1966dynamic,diaz2004optimal}, and multi-stage stochastic optimization \cite{jia2024learning,huang2009value}. Given that this work primarily focuses on network reconfiguration and equity aspects, and that incorporating uncertainties would inevitably cause tractability issues in large systems, the stochastic modeling parts will be reserved for exploration in future studies.
\end{remark}


Despite effectively reducing the total load shedding in the multi-period model, the optimization formulation \eqref{eq:original} lacks consideration of equity and fairness in the operational decisions.
For instance, the optimal solutions may entail frequent load shedding at one load block, while other blocks experiencing much fewer load shedding operations. Additionally, certain loads may include emergency services (e.g., medical centers, fire stations), for which any shutoffs should be avoided as much as possible.
In the following section, we will elaborate on these concerns and introduce a more comprehensive formulation for the equitable network reconfiguration problem.

\section{Equitable Network Reconfiguration Formulation}
\label{sec:equity}
Promoting energy justice and equity \cite{jenkins2016energy,mathieu2023algorithms} has become an emerging topic for modern power grid operations. To achieve this goal, the system operators not only need to strive for fairness among customers in current and future operations, but also need to factor in the vulnerability and priority of customers from different backgrounds.
To this end, our subsequent discussion centers on exploring equity considerations, with the eventual goal of seamlessly incorporating them into the network reconfiguration problem.

\subsection{Individual-level Equity Constraints}
To start with, we formulate constraints on individual load blocks to prevent excessive load shedding. In particular, the total number of load shedding for each load block $\kappa \in \mathcal B$ over the multi-period $t \in \mathcal T$ can be constrained, as given by:
\begin{align} \label{eq:eq_1}
    \sum_{t \in \mathcal T} (1-z_{\kappa,t}) \leq {\alpha}_{\kappa},  \quad \forall \kappa \in \mathcal B
\end{align}
where ${\alpha}_{\kappa} \in \left[0, T\right]$ denotes the load shedding upper limit for block $\kappa$, which can be set up differently for each block based on specific needs.
Note that in the original network reconfiguration problem \eqref{eq:original}, the total load shedding per load block is not constrained at all, and thus ${\alpha}_{\kappa} = T$.

For load blocks hosting emergency services \cite{poudel2018critical} such as hospitals and fire centers where load shedding is definitely undesirable, constraints can be directly enforced to regulate the load shedding decision. 
For ease of exposition, we split the load block set $\mathcal B$ into emergency blocks $\cal M$ and non-emergency blocks $\mathcal{B}\setminus\mathcal{M}$. Hence, the constraints of load shedding for these blocks can be defined as follows:
\begin{subequations} \label{eq:eq_2}
\begin{align} 
     & z_{\kappa,t} = 1, \: \: \forall \kappa \in \mathcal{M},  \: \forall t \in \mathcal T\\
     & 0 \leq z_{\kappa,t} \leq 1, \: \: \forall \kappa \in \mathcal{B}\setminus\mathcal{M},  \: \forall t \in \mathcal T
\end{align}
\end{subequations} 
where emergency load blocks are expected to remain in operation at all times, while other load blocks are allowed to be switched on and off. Note that these constraints are flexible and can be modified for different time intervals $t$ to accommodate specific requirements.

Although load shedding may be necessary at times to maintain power balance and prevent system-wide blackouts, prolonged or frequent load shedding in the same area is clearly undesirable. It diminishes the quality of life, raises safety concerns, and even poses threats to human lives. Therefore, strategic planning of load shedding during emergencies is of critical importance for maintaining both energy equity and public safety.
In order to prevent prolonged load shedding, the following constraint can be implemented to limit the duration of each load shedding event:
\begin{align} \label{eq:eq_3}
    & \sum_{\tau = t}^{t+\Delta t} (1-z_{\kappa,\tau}) \leq \lambda (\Delta t + 1),  \quad \forall \kappa \in {\cal B}
\end{align}
where $\Delta t + 1$ denotes the duration of the considered timeframe and $\lambda$ identifies the maximum ratio of load shedding allowed. For example, for every 24 hours ($\Delta t = 24$), the total duration of load shedding cannot exceed 6 hours ($\lambda = 0.25$). On the other hand, recurrent load shedding within one area is also undesired and should be avoided. This can be achieved by enforcing constraints on the change of load block status. Specifically, the following constraint is presented to limit the frequency of load shedding operations:
\begin{align} \label{eq:eq_4}
    & \sum_{t \in \cal T} \Delta z_{\kappa,t} \leq m, \quad \forall \kappa \in \cal B
\end{align}
where the total change of load block status is limited to $m$ times. The change of load shedding status on block $\kappa$ at time $t$ is defined by a binary decision variable:
\begin{align} \label{eq:eq_5}
    \Delta z_{\kappa,t} = |z_{\kappa,t} - z_{\kappa,t-1}|, \quad \forall t = 2, \ldots, T. 
\end{align}
If $\Delta z_{\kappa,t} = 1$, it indicates the load block status is either changed from 1 to 0, or from 0 to 1. On the other hand, $\Delta z_{\kappa,t} = 0$ means the load block status remains the same from time $t-1$ to $t$. 
The absolute value function in \eqref{eq:eq_5} can be handled with epigraph reformulation using the following two constraints:
\begin{subequations} \label{eq:eq_6}
\begin{align}
    & \Delta z_{\kappa,t} \geq z_{\kappa,t} - z_{\kappa,t-1}, \quad \forall t = 2, \ldots, T \label{eq:eq_6_a}\\
    & \Delta z_{\kappa,t} \geq z_{\kappa,t-1} - z_{\kappa,t}, \quad \forall t = 2, \ldots, T \label{eq:eq_6_b}
\end{align}
\end{subequations}
In particular, when load block status changes $z_{\kappa,t} \neq z_{\kappa,t-1}$, \eqref{eq:eq_6_a} and \eqref{eq:eq_6_b} jointly guarantee $\Delta z_{\kappa,t} \geq \pm 1$ and thus $\Delta z_{\kappa,t} = 1$. On the other hand, if a load shedding is not performed $z_{\kappa,t} = z_{\kappa,t-1}$, binary variable $\Delta z_{\kappa,t} \geq 0$ and thus assume a value of 0 or 1. Nonetheless, under either case, constraint \eqref{eq:eq_4} will ensure that the total change of load block status is restricted to a maximum of $m$ times.

\subsection{Equity Constraints Through Comparison}
To secure fairness of the shutoff decisions, it is important to guarantee that none of the local load shedding is disproportionately high as compared to the total load shedding operations. Given a predetermined threshold $\psi_{\kappa}$ for each load block $\kappa$, the following constraint is presented:
\begin{align} \label{eq:eq_7}
    \sum_{t\in \cal T} (1-z_{\kappa,t}) \leq \psi_{\kappa} \sum_{t\in \cal T} \sum_{\kappa \in \cal B} (1-z_{\kappa,t}),  \: \forall \kappa \in {\cal B}
\end{align}
where the total number of load shedding of a selected block $\kappa$ cannot exceed a fixed percentage (e.g., $\psi_{\kappa} = 20\%$) of the system-wide total load shedding. Similarly, the comparison can also be executed for any pair of load blocks $(\kappa,\nu)$:
\begin{align} \label{eq:eq_8}
     \frac{\sum_{t\in \cal T} (1-z_{\kappa,t})}{\sum_{t\in \cal T} (1-z_{\nu,t})} \leq \beta_{\kappa\nu}, \quad \forall \kappa, \nu \in {\cal B}.
\end{align}
Here the parameter $\beta_{\kappa\nu}$ denotes the upper bound on the ratio of total load shedding at block $\kappa$ and $\nu$. The linear constraint helps prevent the load shedding at one block to significantly deviate from one another. 
Lastly, for equitable load shedding, it also makes sense to factor in the vulnerability of customers. The vulnerability of a load block may be influenced by factors like community location, average income, infrastructure quality, and emergency preparedness. For simplicity, we assume that all these factors have been accounted for, and each load block at time $t$ is assigned a comprehensive vulnerability index \cite{taylor2023managing} $v_{\kappa,t}$. Therefore, the decision-making problem should aim for an optimal strategy that simultaneously minimizes the overall vulnerability against emergency events.
Notably, the vulnerability index $v_{\kappa,t}$ can be informative for the design of parameter $\beta_{\kappa\nu}$ in \eqref{eq:eq_8}. In general, an equitable shutoff strategy should anticipate less load shedding at the load block with a relatively high vulnerability index. Therefore, the ratio $\frac{\sum_{t\in \cal T} (1-z_{\kappa,t})}{\sum_{t\in \cal T} (1-z_{\nu,t})}$ should roughly scale with the inverse of the vulnerability of the associated load blocks, which suggests the design of parameter $\beta_{\kappa\nu}$ to typically follow:
\begin{align}
    \beta_{\kappa\nu}  \propto \frac{\sum_{t\in \cal T} v_{\nu,t}}{\sum_{t\in \cal T} v_{\kappa,t}}, \quad \forall \kappa, \nu.
\end{align}
Broadly speaking, the selection of parameter $\beta_{\kappa\nu}$ should be inclined towards supporting more vulnerable communities to enhance social equity.

\begin{itemize}[label={$\bullet$}]
    \item \textit{Equitable Network Reconfiguration Problem} 
\end{itemize}

Based on all the discussions thus far, the equitable network reconfiguration problem can be cast as the following optimization formulation:
\begin{subequations} \label{eq:equity}
\begin{align}
\min \quad & \sum_{t\in \mathcal T} \sum_{\kappa \in \mathcal B} {P^{d}_{\kappa,t} (1-z_{\kappa,t})} + \rho\sum_{t\in \mathcal T} \sum_{\kappa \in \mathcal B} {v_{\kappa,t} (1-z_{\kappa,t})} \label{eq:E_a}\\
\textrm{s.t.} \quad &  \eqref{eq:pf_1} - \eqref{eq:con_topology5}   \label{eq:E_b}\\
 &  \eqref{eq:eq_1} - \eqref{eq:eq_8}   \label{eq:E_c}
\end{align}
\end{subequations}
which incorporates diverse equity constraints in addition to the original network reconfiguration problem. The vulnerability index is only added to the objective when the corresponding load block is switched off. The coefficient $\rho$ balance between total load shedding costs and total vulnerability costs. The constraints include those from the original network reconfiguration problem \eqref{eq:original} and the equity constraints introduced in this section. In contrast, the network reconfiguration problem \eqref{eq:equity} provides a more equitable and reliable shutoff strategy, by considering social vulnerability, and energy equity. Unlike the strict implementation of system operating constraints \eqref{eq:pf_1} - \eqref{eq:con_topology5}, the equity constraints \eqref{eq:eq_1} - \eqref{eq:eq_8} offer a high degree of flexibility, allowing for convenient modifications to align with diverse system needs.

\section{Numerical Results}
\label{sec:num}

In this section, we present the numerical results to validate the network reconfiguration performance using a modified version of the IEEE 13-bus feeder system and a more realistic larger-sized SmartDS test system. 
Simulations are performed on a regular laptop with Intel\textsuperscript{\textregistered} CPU @ 2.60 GHz and 16 GB of RAM. The network reconfiguration problem is implemented with PowerModelsONM package \cite{fobes2022} using Julia, and the mixed-integer optimization problem is solved using the CPLEX solver, with up to 12 threads and a tolerance gap of $0.01\%$.
The nominal wildfire risk value $r_{\kappa}$ and vulnerability index $v_{\kappa}$ for each load block are adopted from datasets provided in \cite{taylor2023managing}. Specifically, in the SmartDS system, the Wildland Fire Potential Index \cite{WFPI} is employed to generate the wildfire risk data, and CDC SVI dataset is utilized to quantify the social vulnerability. Through the numerical validation, we assume that the social vulnerability parameter $v_{\kappa}$ for each individual load block remains consistent over the entire horizon. Meanwhile, the wildfire risks are time-dependent and will be specified in detail later. 

\vspace{-0.3cm}

\subsection{IEEE 13-bus system}

\begin{figure}[t!]
\centering
\includegraphics[trim=6.5cm 5cm 0cm 4cm,clip=true,totalheight=0.18\textheight]{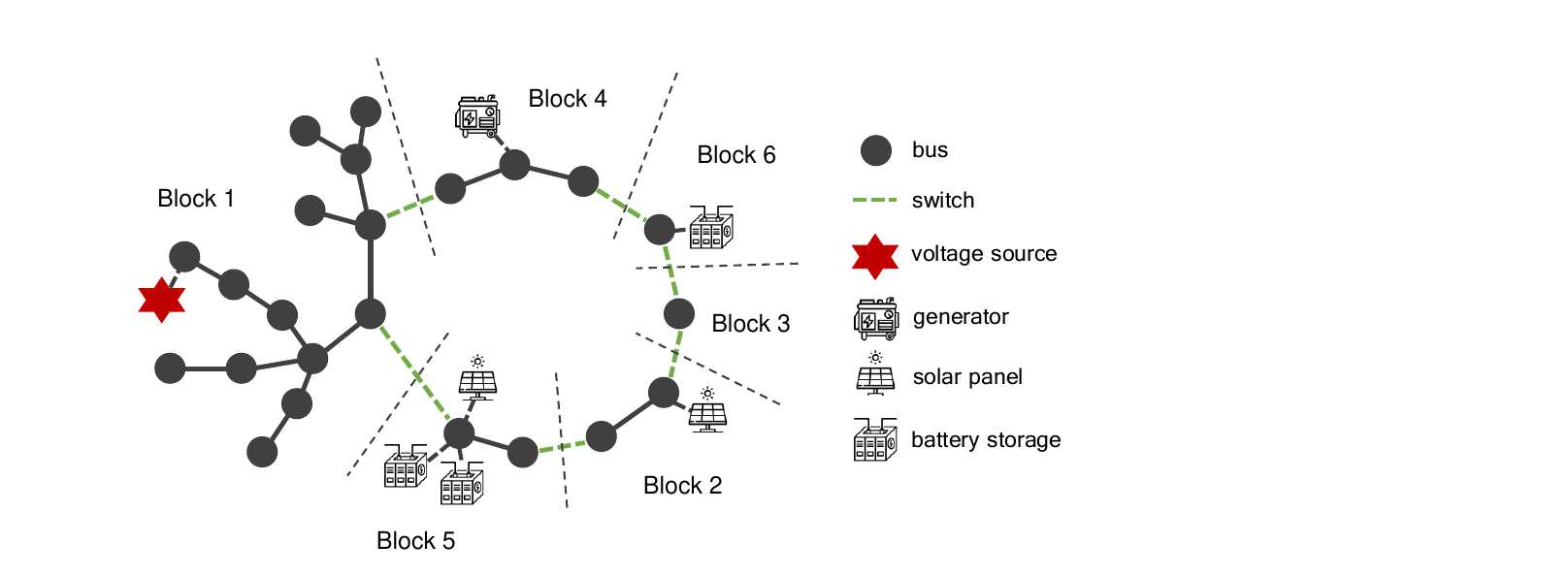}
\caption{The modified test case from IEEE 13 bus system.}
\label{fig:13_bus_new}
\vspace{-0.5cm}
\end{figure}


\begin{figure}[t!]
\centering
\includegraphics[trim=0cm 0cm 0cm 0cm,clip=true,totalheight=0.14\textheight]{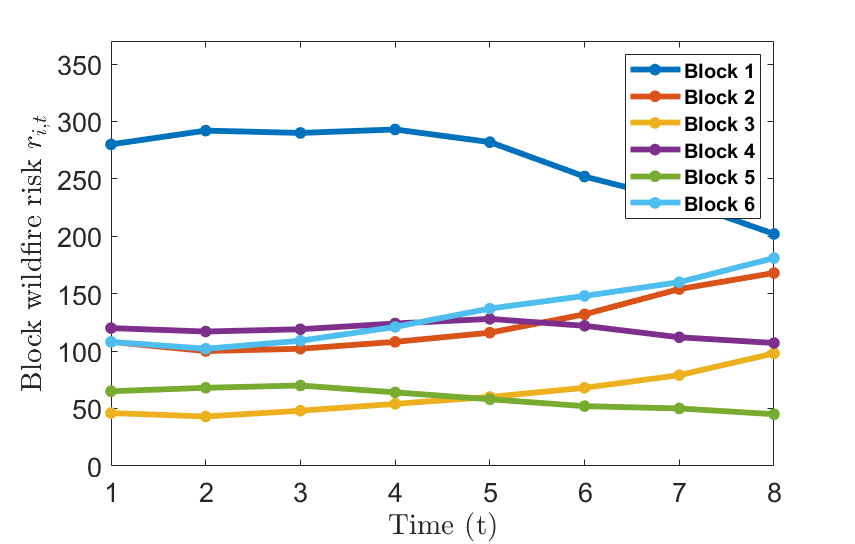}
\caption{The wildfire risk $r_{\kappa,t}$ values for different load blocks.}
\label{fig:risk_13_bus}
\vspace{-0.5cm}
\end{figure}

We first consider a modified version of the IEEE 13-bus distribution feeder case. The networked system is demonstrated in Fig. \ref{fig:13_bus_new}, which consists of 23 buses (6 load blocks), and 6 distributed energy resources (DERs). The DERs include 1 traditional generator, 2 photovoltaic (PV) units, and 3 battery storage devices. In addition, there are a total of 6 switches among load blocks to allow for flexible network switching.

\begin{table*}[t!] 
  \centering
  \caption{Block connection status over the rolling horizon for the 13-bus system (block 6 is reserved for emergency services)}
\begin{tabular}{ 
  |P{2.4cm}||P{1.3cm}|P{1.3cm}|P{1.3cm}|P{1.3cm}|P{1.3cm}|P{1.3cm}|P{1.3cm}|P{1.3cm}|P{1.3cm}| }
 \hline
  Time & t = 1 & t = 2 & t = 3 & t = 4 & t = 5 & t = 6 & t = 7 & t = 8 \\
 \hline
 Block 1  & 0  & 1 & 1 & 1 & 1 & 1 & 1 & 1 \\
  \hline
 Block 2  & 1 & 1  & 0 & 0 & 0 & 0 & 0 & 0  \\
  \hline
 Block 3  & 1  & 0  & 0 & 0 & 0 & 0 & 0 & 1 \\
  \hline
 Block 4 & 0  & 0  & 1 & 1 & 1 & 1 & 1 & 0 \\
  \hline
 Block 5 & 1  & 0  & 0 & 0 & 0 & 0 & 1 & 1 \\
  \hline
 Block 6 & 1  & 1  & 1 & 1 & 1 & 1 & 1 & 1 \\
  \hline  
  \end{tabular} \label{tab:1}
\end{table*}

\subsubsection{Experiment Setup}
The equitable network reconfiguration problem is tested on the system, spanning an 8-hour time horizon during the day ($T=8$). The system load profile in percentage over the rolling horizon is included in the source code of PowerModelsONM.jl. 
The nominal $P^{d}_{\kappa}$ values for block 1 to block 6 are $\{2453, 185, 0, 1013, 25, 200\}$ kW, respectively. Meanwhile, the comprehensive vulnerability indices for these blocks are set as $\{2,9,2,4,6,3\}$.
To provide intra-day variations for evaluating the rolling horizon problem, we vary the wildfire risks for different blocks from their nominal values. Specifically, the data is constructed to mimic a trend of wildfire risks geographically moving from left to right for the system in Fig. \ref{fig:13_bus_new}. When $t$ increases, the wildfire risks $r_{\kappa,t}$ for blocks 1, 4, and 5 gradually decrease while the risks for other blocks become larger, which is shown in Fig. \ref{fig:risk_13_bus}.
The safety threshold for wildfire ignition risk in \eqref{eq:con_fire} is set to be $\epsilon = 0.5$. 
To contain the total wildfire risk, selected load blocks need to be shut off for safety concerns. For each time interval $t$, we allow a maximum of 3 load blocks to be shut off at the same time.

\subsubsection{Simulation Results and Analysis}
Solving the network reconfiguration problem under these settings provides the optimal connection status for each load block over the rolling horizon, as shown in TABLE \ref{tab:1}. Among all the load blocks, block 6 is reserved for emergency services and thus cannot be switched off at any time. This can be easily verified from the results, as $z_{6,t} = 1, \forall t \in \cal T$. The remaining load blocks have the flexibility to be either switched on ($z_{\kappa,t} = 1$) or switched off ($z_{\kappa,t} = 0$). To avoid excessive load shedding within individual load blocks, we set $\sum_{t \in \mathcal T} (1-z_{\kappa,t}) \leq 6$ for these non-emergency load blocks (see equation \eqref{eq:eq_1}), as also evidenced by the results in TABLE \ref{tab:1}. 
In addition, to prevent frequent load shedding in individual load blocks, we also include the constraint \eqref{eq:eq_4} where a parameter $m = 2$ is specified. Of all the load blocks, blocks 3, 4, and 5 experience a change of connection status twice while other blocks experience less frequent ($\leq 1$) status changes.

Notably, all non-emergency load blocks are scheduled to perform load shedding operations at least once during the considered rolling horizon. Compared to other non-emergency load blocks, blocks 1 and 4 experience relatively shorter duration of load shedding. This is because the demand of these two blocks is significantly higher than the rest of the system, and thus the shutoff is avoided as much as possible. By and large, the wildfire risks for block 1 and 4 are higher in the early time periods ($t = 1$ and $t = 2$), explaining why load shedding operations are predominantly performed during these intervals. On the other hand, load shedding operations on blocks 2, 3, and 5 mainly center around mid to later time periods, aligning with their relatively large wildfire risks during those intervals. This implies that when load shedding is unavoidable for a certain block, there is a tendency in the decision-making to schedule that shutoff when the corresponding risk is larger.

\begin{figure*}[ht]
\centering
   \subfloat[]{
      \includegraphics[trim=0cm 0cm 0cm 0cm,clip=true, width=0.31\textwidth]{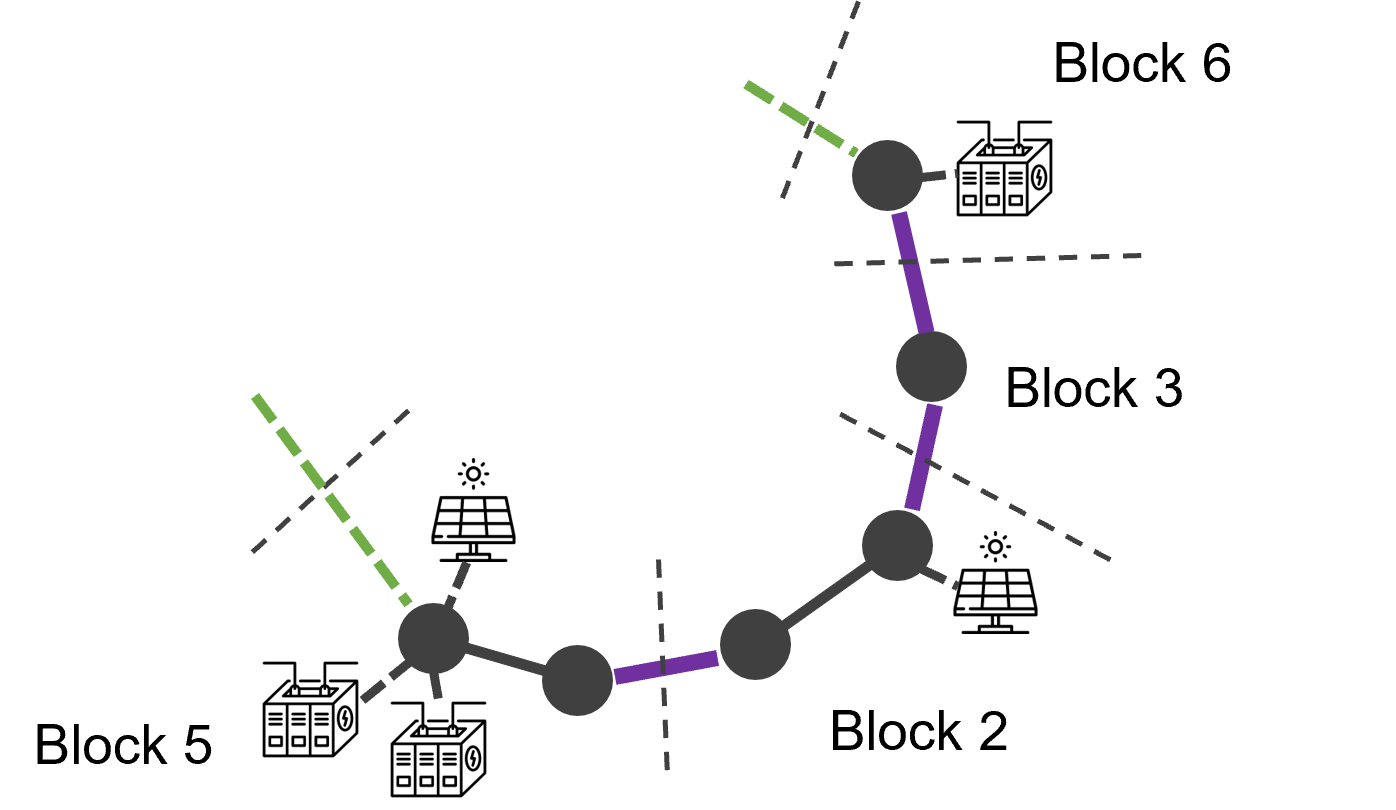}}
\hspace{\fill}
   \subfloat[]{
      \includegraphics[trim=0cm 0cm 0cm 0cm,clip=true, width=0.31\textwidth]{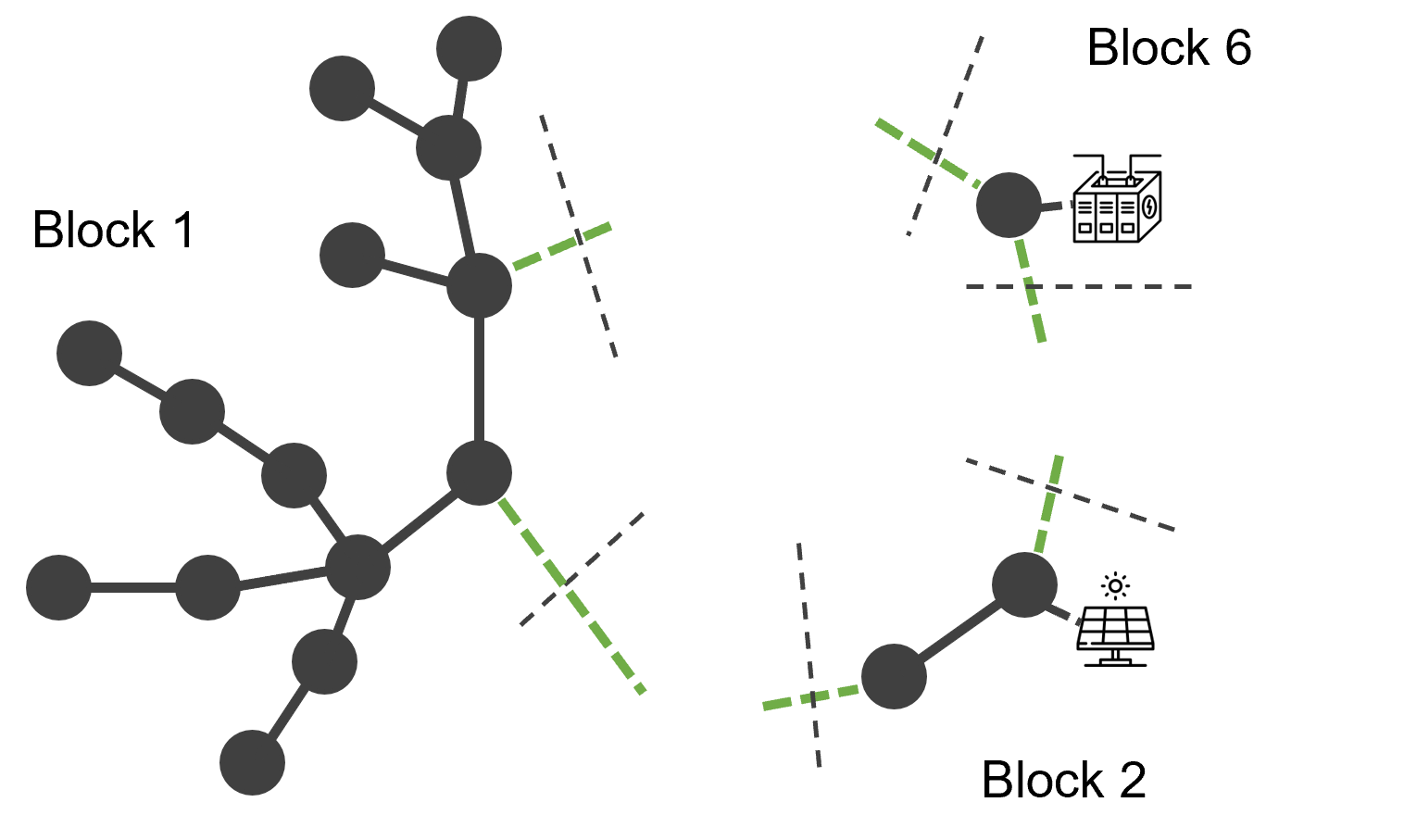}}
\hspace{\fill}
   \subfloat[]{
      \includegraphics[trim=0cm 0cm 0cm 0cm,clip=true, width=0.31\textwidth]{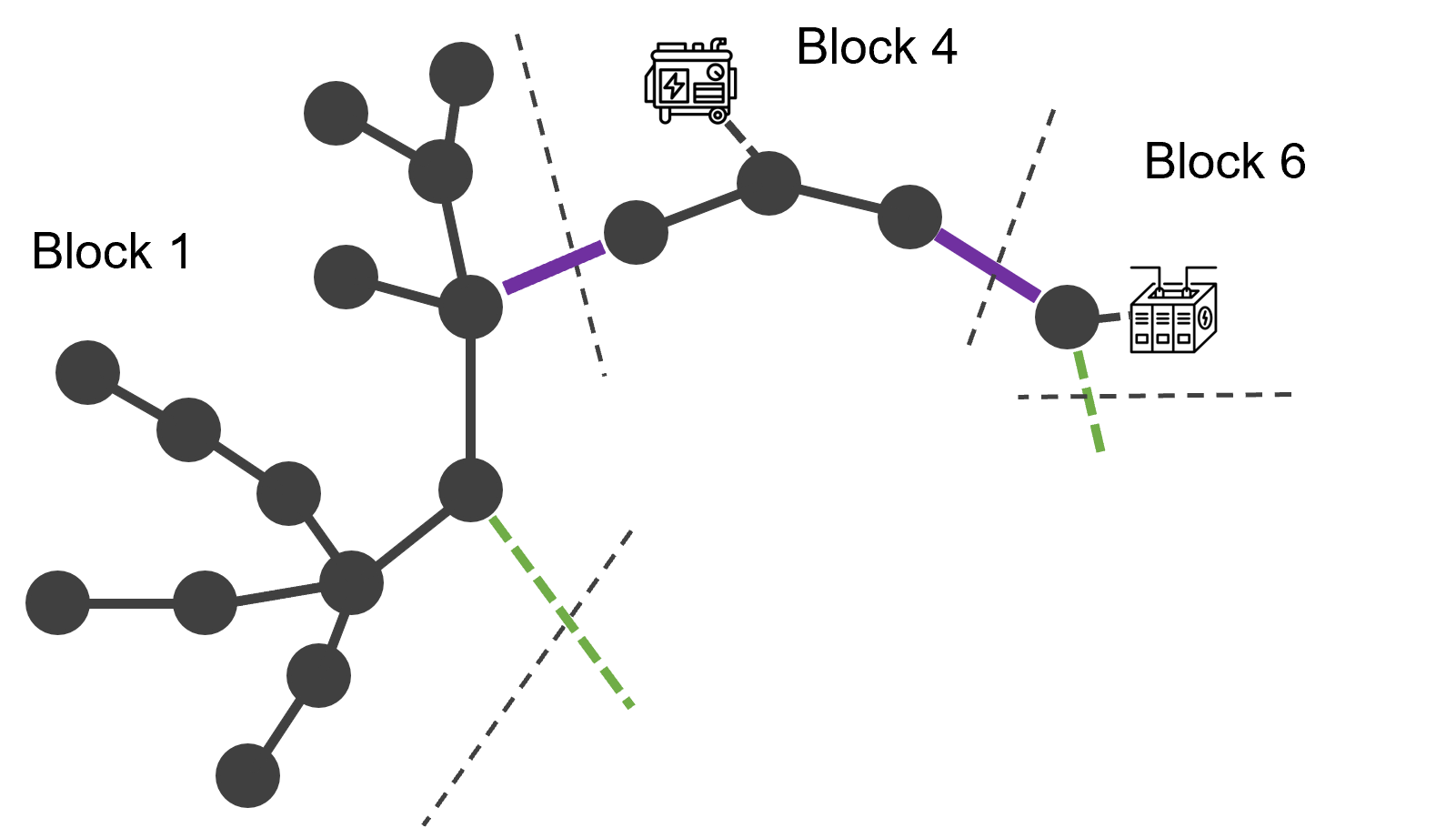}}\\
\caption{The network topology of the 13-bus test case after reconfiguration at time (a) $t = 1$; (b) $t = 2$; and (c) $t = 3$.}
\label{fig:13_bus_RH}
\vspace{-0.4cm}
\end{figure*}

\begin{table}[t!] 
  \centering
  \caption{Block connection status over the rolling horizon for the 13-bus system with parameters $\psi = 100\%$ and $\epsilon$ = 0.8.}
\begin{tabular}{ 
  |P{1.1cm}||P{0.4cm}|P{0.4cm}|P{0.4cm}|P{0.4cm}|P{0.4cm}|P{0.4cm}|P{0.4cm}|P{0.4cm}|P{0.4cm}| }
 \hline
 \backslashbox[1.5cm]{Block}{Time}
   & 1 & 2 & 3 & 4 & 5 & 6 & 7 & 8 \\
 \hline
 Block 1  & 1 & 1 & 1 & 1 & 1 & 1 & 1 & 1 \\
  \hline
 Block 2  & 1 & 1 & 0 & 0 & 0 & 0 & 0 & 0  \\
  \hline
 Block 3  & 1 & 0 & 0 & 0 & 0 & 0 & 0 & 1 \\
  \hline
 Block 4  & 0 & 0 & 1 & 1 & 1 & 1 & 1 & 1 \\
  \hline
 Block 5  & 0 & 1 & 1 & 1 & 1 & 1 & 1 & 1 \\
  \hline
 Block 6  & 1 & 1 & 1 & 1 & 1 & 1 & 1 & 1 \\
  \hline  
  \end{tabular} \label{tab:2}
\end{table}

\begin{table}[t!] 
  \centering
  \caption{Block connection status over the rolling horizon for the 13-bus system with parameters $\psi = 35\%$ and $\epsilon$ = 0.8.}
\begin{tabular}{ 
  |P{1.1cm}||P{0.4cm}|P{0.4cm}|P{0.4cm}|P{0.4cm}|P{0.4cm}|P{0.4cm}|P{0.4cm}|P{0.4cm}|P{0.4cm}| }
 \hline
 \backslashbox[1.5cm]{Block}{Time}
   & 1 & 2 & 3 & 4 & 5 & 6 & 7 & 8 \\
 \hline
 Block 1  & 1 & 1 & 1 & 1 & 1 & 1 & 1 & 1 \\
  \hline
 Block 2  & 1 & 1 & 0 & 0 & 0 & 0 & 0 & 1  \\
  \hline
 Block 3  & 1 & 1 & 1 & 0 & 0 & 0 & 0 & 0 \\
  \hline
 Block 4  & 0 & 0 & 1 & 1 & 1 & 1 & 1 & 0 \\
  \hline
 Block 5  & 0 & 0 & 0 & 1 & 1 & 1 & 1 & 1 \\
  \hline
 Block 6  & 1 & 1 & 1 & 1 & 1 & 1 & 1 & 1 \\
  \hline  
  \end{tabular} \label{tab:3}
  \vspace{-0.5cm}
\end{table}

The solution to the network reconfiguration problem may lead to variations in the system topology during different time intervals. 
Fig. \ref{fig:13_bus_RH} illustrates the network connection status during the first three time periods. After certain blocks are shut off, the rest of the network forms either a single or a few individually-operated microgrids. Each microgrid is equipped with a grid-forming inverter, to which all local energized load blocks are connected. The demand for these microgrids is supplied by all available DERs within the sub-network following network reconfiguration.

\subsubsection{Sensitivity Validation of Parameter $\epsilon$}
The setup of the optimization problem has a significant impact on the number of load blocks to be switched off. 
Specifically, the number of buses to be switched is heavily affected by the safety threshold $\epsilon$ specified in constraint \eqref{eq:con_fire}. In general, a smaller $\epsilon$ provides more security guarantees at the cost of more shutoffs. 
To validate this, we increase the threshold $\epsilon$ from $0.5$ to $0.8$, which leads to the optimal network reconfiguration as shown in TABLE \ref{tab:2}. 
Although the upper limit for shutoffs at each time interval remains to be $3$, the actual number of blocks shut off is decreased to $1$ or $2$. This is because the constraint on wildfire risk is alleviated with a larger $\epsilon$, and thus fewer shutoffs are required. Note that both blocks 2 and 3 are shut off $6$ times, contributing to $40\%$ of the total 15 switching operations across the system. To ensure fairness for these two blocks as compared to other blocks, we uniformly impose constraint \eqref{eq:eq_7} for all non-emergency load blocks equally where $\psi_{\kappa}$ is set to be $35\%$ and the results are shown in TABLE \ref{tab:3}.
The major difference in optimal decisions after introducing this additional equity constraint is that part of load shedding operations scheduled for blocks 2 and 3 are now undertaken by blocks 4 and 5 to promote equity, while all other constraints earlier remain satisfied. Of course, the choice of equity constraints can be made strategically according to different system configurations and the duration of the rolling horizon. These considerations will be investigated using a more realistic distribution system model next.

\subsection{Smart-DS System}

We also validate the numerical performance of the proposed algorithm on the Smart-DS test system \cite{krishnan2017smart}. The Smart-DS stands for Synthetic Models for Advanced, Realistic Testing: Distribution Systems and Scenarios, which is a realistic dataset representing distribution models in the San Francisco area, in California. The data parsing was performed with the PowerModelsONM package \cite{fobes2022}. The test feeder has been slightly reduced, and the modified system consists of 846 buses (57 load blocks),  784 lines, 124 generators, 278 loads, and 132 switches. In addition, it comes with 123 photovoltaic (PV) units and 308 energy storage units.
The time-series loads are generated using a realistic hourly load profile throughout the day.
In addition, we obtain the nominal wildfire risks using Wildland Fire Potential Index (WFPI) \cite{WFPI} and then construct the complete time-series wildfire risks based on hourly temperature data mapped to geographic coordinates.
Lastly, the quantification of each load block's vulnerability to power outages is obtained using the CDC Social Vulnerability Index (SVI). The census tract-based SVI data is overlaid with the Smart-DS feeder region, to generate the vulnerability values. These values represent the long-term vulnerability of each load block to power outages, and thus we keep them consistent over the entire rolling horizon.


\begin{table*}[t!] 
  \centering
  \caption{Load blocks scheduled for shutoff in Smart-DS under the original network reconfiguration problem}
\begin{tabular}{ 
  |P{1.1cm}||P{0.25cm}|P{0.25cm}|P{0.25cm}|P{0.25cm}|P{0.25cm}|P{0.25cm}|P{0.25cm}|P{0.25cm}|P{0.25cm}|P{0.25cm}|P{0.25cm}|P{0.25cm}|P{0.25cm}|P{0.25cm}|P{0.25cm}|P{0.25cm}|P{0.25cm}| P{0.25cm}|P{0.25cm}|P{0.25cm}|P{0.25cm}| P{0.25cm}|P{0.25cm}|P{0.25cm}|P{0.25cm}| P{0.25cm}|P{0.25cm}|P{0.25cm}|P{0.25cm}| }
 \hline
  Time & 1 & 2 & 3 & 4 & 5 & 6 & 7 & 8 & 9 & 10 & 11 & 12 & 13 & 14 & 15 & 16 & 17 & 18 & 19 & 20 & 21 & 22 & 23 & 24\\
 \hhline{|=========================|}
 Block \#  & 9 & 10 & 10 & 3 & 10 & 2 & 9 & 10 & 10 & 4 & 10 & 9 & 10 & 9 & 10 & 10 & 10 & 10 & 10 & 4 & 10 & 10 & 10 & 3 \\
  \hline
 Block \#   & 10 & 24 & 19 & 10 & 21 & 9 & 10 & 24 & 24 & 10 & 24 & 10 & 24 & 10 & 22 & 12 & 24 & 12 & 24 & 10 & 23 & 24 & 24 & 10  \\
  \hline
 Block \#   & 24 & 40 & 24 & 24 & 23 & 10 & 24 & 36 & 51 & 24 & 41 & 14 & 29 & 24 & 24 & 24 & 29 & 23 & 40 & 24 & 24 & 54 & 51 & 24 \\
  \hline
 Block \#   & 53 & 52 & 41 & 40 & 24 & 24 & 44 & 53 & 54 & 28 & 54 & 24 & 49 & 48 & 44 & 36 & 40 & 24 & 54 & 40 & 32 & 55 & 54 & 54 \\
  \hline
 Block \#   & 54 & 54 & 54 & 54 & 54 & 54 & 55 & 54 & 55 & 54 & 55 & 54 & 54 & 54 & 54 & 54 & 54 & 54 & 56 & 54 & 54 & 56 & 55 & 56 \\
  \hline
  \end{tabular} \label{tab:4}
\end{table*}

\begin{table*}[t!] 
  \centering
  \caption{Load blocks scheduled for shutoff in Smart-DS under the equitable network reconfiguration problem}
\begin{tabular}{ 
  |P{1.1cm}||P{0.25cm}|P{0.25cm}|P{0.25cm}|P{0.25cm}|P{0.25cm}|P{0.25cm}|P{0.25cm}|P{0.25cm}|P{0.25cm}|P{0.25cm}|P{0.25cm}|P{0.25cm}|P{0.25cm}|P{0.25cm}|P{0.25cm}|P{0.25cm}|P{0.25cm}| P{0.25cm}|P{0.25cm}|P{0.25cm}|P{0.25cm}| P{0.25cm}|P{0.25cm}|P{0.25cm}|P{0.25cm}| P{0.25cm}|P{0.25cm}|P{0.25cm}|P{0.25cm}| }
 \hline
  Time & 1 & 2 & 3 & 4 & 5 & 6 & 7 & 8 & 9 & 10 & 11 & 12 & 13 & 14 & 15 & 16 & 17 & 18 & 19 & 20 & 21 & 22 & 23 & 24\\
 \hhline{|=========================|}
 Block \#   & 10 & 25 & 25 & 25 & 24 & 24 & 24 & 23 & 24 & 36 & 32 & 11 & 11 & 11 & 20 & 15 & 12 & 12 & 12 & 7 & 14 & 19 & 19 & 19 \\
  \hline
 Block \#   & 25 & 30 & 30 & 30 & 25 & 30 & 30 & 24 & 34 & 50 & 52 & 29 & 16 & 16 & 28 & 32 & 15 & 14 & 14 & 14 & 19 & 21 & 21 & 21  \\
  \hline
 Block \#   & 37 & 37 & 37 & 31 & 30 & 48 & 36 & 36 & 36 & 52 & 53 & 32 & 29 & 29 & 29 & 40 & 40 & 40 & 19 & 19 & 21 & 45 & 45 & 45 \\
  \hline
 Block \#   & 47 & 47 & 47 & 37 & 37 & 49 & 48 & 52 & 52 & 55 & 56 & 52 & 32 & 32 & 32 & 41 & 41 & 41 & 41 & 21 & 33 & 46 & 46 & 46 \\
  \hline
 Block \#   & \slash & 49 & 49 & 49 & 49 & 57 & 57 & 57 & 56 & 56 & \slash & 56 & 52 & 41 & 41 & 43 & 43 & 43 & 43 & 27 & 38 & 47 & 47 & 47 \\
  \hline
  \end{tabular} \label{tab:5}
\end{table*}

\begin{figure}[t!]
\centering
\includegraphics[trim=0cm 0cm 0cm 0cm,clip=true,totalheight=0.13\textheight]{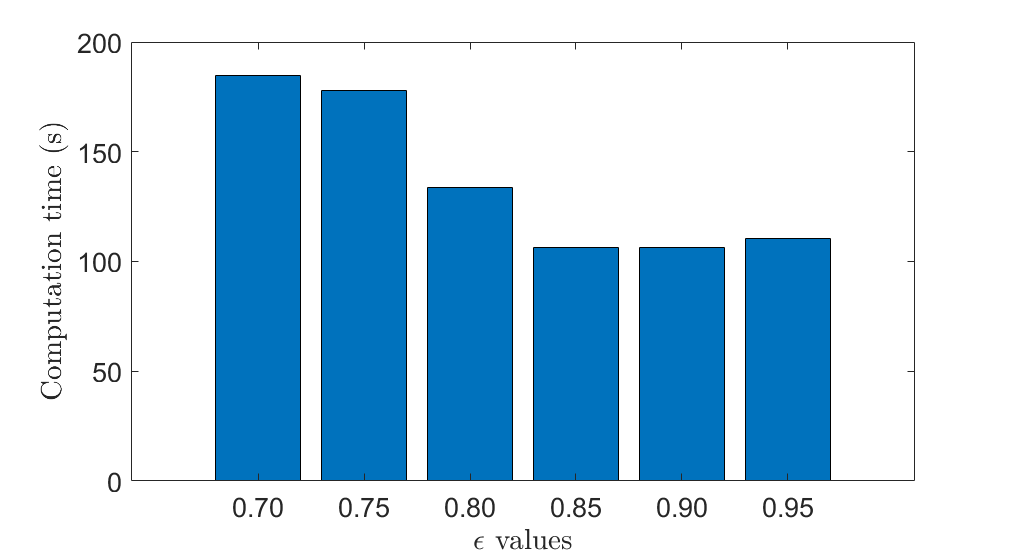}
\caption{Comparison of computation time under different $\epsilon$ values.}
\label{fig:DS_time}
\vspace{-0.5cm}
\end{figure}

\begin{figure}[t!]
\centering
\includegraphics[trim=0cm 0cm 0cm 0cm,clip=true,totalheight=0.14\textheight]{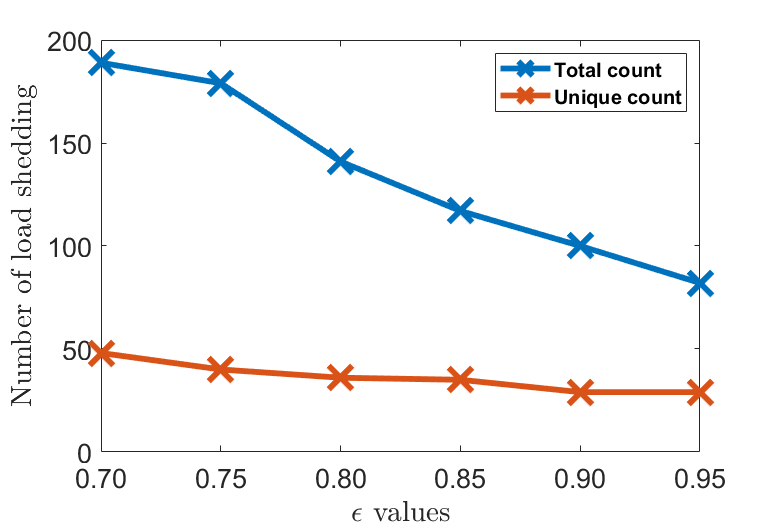}
\caption{Comparison of load shedding under different $\epsilon$ values.}
\label{fig:DS_number}
\vspace{-0.5cm}
\end{figure}

\textit{Test Case 1.} We first test the network reconfiguration problem using hourly data for 24 consecutive hours ($t \in \{1,2,\cdots,24\}$). In particular, we compare the results for the original network reconfiguration problem \eqref{eq:original} and the problem \eqref{eq:equity} with equity formulations. For both problems, we set the wildfire safety threshold $\epsilon = 90\%$ and allow a maximum of 5 load blocks to be shut off during every time interval. For the equitable network reconfiguration problem, we further limit the total shutoffs for each load block to no more than 6 times throughout the day. Load blocks 1-5 are configured to be emergency blocks, and for the remaining blocks, the number of shutoffs is restricted to a maximum of 2 times.
The average runtime for the original network reconfiguration problem is 1743 seconds, whereas the problem with equitable formulations only takes 779 seconds to solve on average. One possible cause for the variation is that the equitable formulations impose additional constraints on the binary switching variables, effectively narrowing down the search space and accelerating the computation. Enhanced computational capabilities provided by high-performance computing resources, together with optimized parameter settings can further facilitate its real-time implementation.



\begin{figure*}[t!]
\centering 
   \subfloat[$\beta = \infty$]{
      \includegraphics[trim=0cm 0cm 0cm 0cm,clip=true, width=0.25\textwidth]{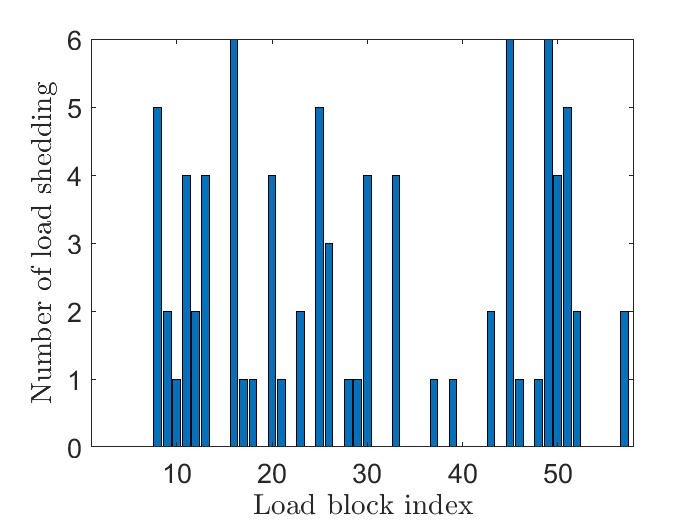} \label{fig:DS_3_1}} 
\hspace{-0.6cm}
   \subfloat[$\beta = 6$]{
      \includegraphics[trim=0cm 0cm 0cm 0cm,clip=true, width=0.25\textwidth]{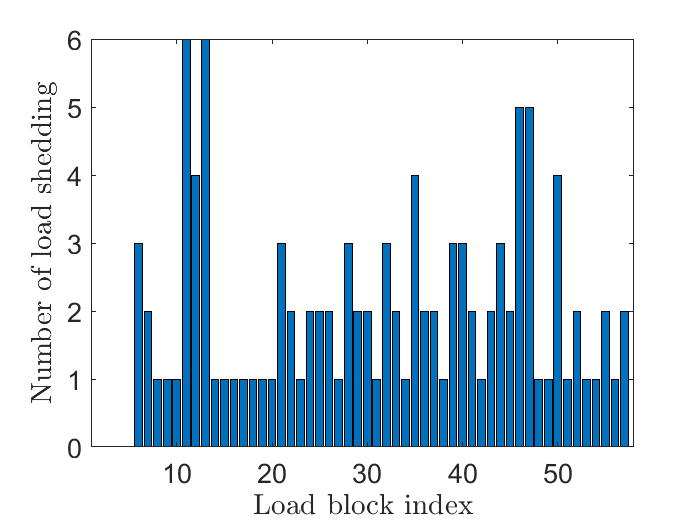} \label{fig:DS_3_2}}
\hspace{-0.6cm}
   \subfloat[$\beta = 4$]{
      \includegraphics[trim=0cm 0cm 0cm 0cm,clip=true, width=0.25\textwidth]{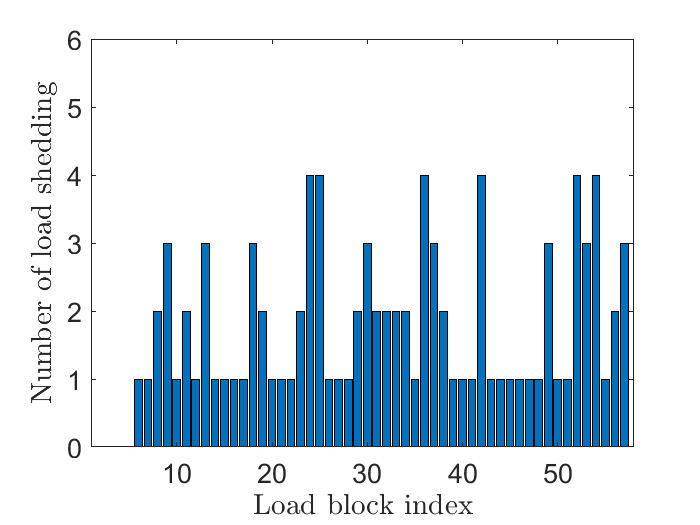} \label{fig:DS_3_3} }
\hspace{-0.6cm}
   \subfloat[$\beta = 2$]{
      \includegraphics[trim=0cm 0cm 0cm 0cm,clip=true, width=0.25\textwidth]{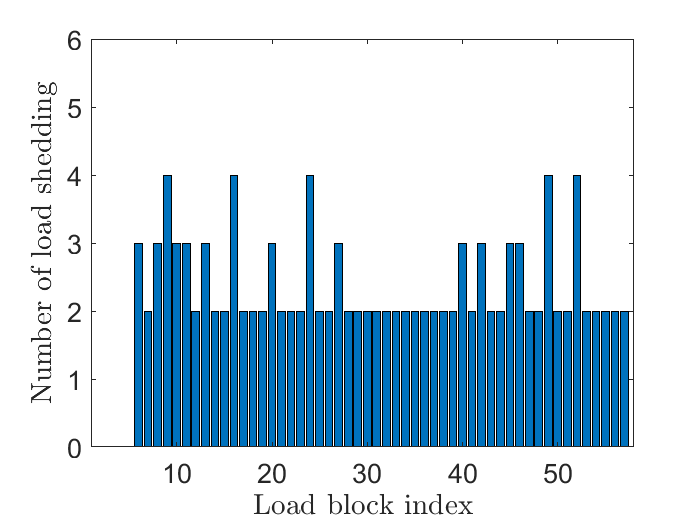} \label{fig:DS_3_4} }
\caption{The comparison of the number of total load shedding at each load block under different $\beta$ values.}
\label{fig:DS_3}
\vspace{-0.4cm}
\end{figure*}

The shutoff decisions for both problems are illustrated in TABLE \ref{tab:4} and TABLE \ref{tab:5}, respectively.
The original network reconfiguration problem involves load shedding operations across 25 load blocks within the 24-hour timeframe. Among these load blocks, certain blocks (e.g., 10, 24, 54) experience more frequent load shedding operations.
In contrast, the equitable network reconfiguration problem involves a total of 35 load blocks to participate in the shutoffs over the entire horizon. Most of these load blocks are scheduled for shutoff for 3-6 consecutive hours while others experience shorter duration of shutoffs. Overall, the equitable formulation offers an improved shutoff strategy, promoting fairness among load blocks and preventing situations where certain load blocks are consistently shut off.
Compared with the original reconfiguration problem without equity consideration, the equitable shutoff decisions from problem \eqref{eq:equity} can be affected by several factors. For instance, a higher coefficient $\rho$ would prioritize the vulnerability objective over total load shedding, potentially resulting in an increased system-wide load shedding. On the other hand, modifications of equity constraint parameters can also visibly affect the shutoff strategies. For example, a reduced wildfire safety threshold $\epsilon$ can provide extra system security at the cost of more shutoffs. Meanwhile, reducing the limit of shutoff duration would likely include more load blocks for shutoffs such that the total wildfire risk can be contained. 
Typically, under a fixed weighting factor $\rho$, the implementation of more restrictive equity constraints is often associated with higher objective costs, but it can provide a more balanced and equitable shutoff strategies.

\textit{Test Case 2.} This part mainly investigates how the wildfire risk threshold affects the shutoff decisions in the equitable network reconfiguration problem. To achieve faster convergence of results, the duration of the horizon is reduced to 12 hours in this test case. We vary the wildfire risk safety threshold $\epsilon$ from $70\%$ to $95\%$ while keeping all other parameters unchanged. As a reduced $\epsilon$ is often associated with more shutoffs, a small upper bound $\overline{k}_{\text{bl}}$ in \eqref{eq:switch_limit_block} could render infeasible solutions. To prevent this, we uniformly increase $\overline{k}_{\text{bl}}$ to 16 for all scenarios in this test. By varying the $\epsilon$ values, our primary objective is to assess the impact of this threshold on the number of shutoffs. Specifically, we compare the computation time, the total number of load shedding as well as the number of unique load blocks involved in load shedding under different $\epsilon$ values.

The results of the comparison are shown in Fig.~\ref{fig:DS_time} and Fig.~\ref{fig:DS_number}.
The computation time for all test scenarios is approximately 150 seconds, with slightly longer time observed for smaller $\epsilon$ values. This could be attributed to the fact that a smaller threshold typically involves more shutoff decisions to make, as also evidenced by Fig.~\ref{fig:DS_number}. As $\epsilon$ is gradually decreased from 0.95 to 0.70, the total number of shutoffs over the 12-hour horizon increases from 82 to 189.  Meanwhile, the unique number of load blocks participating in shutoffs increases from only 29 to 48, which almost reaches the number of all non-emergency load blocks in the system.
It indicates that when extensive shutoffs are necessary, the majority of load blocks will actively participate in reducing the risk instead of letting a few load blocks take on the entire task. This observation from the results very well aligns with our initial vision of ensuring fairness and equity in the network reconfiguration problem.

\textit{Test Case 3.} In this part, our primary focus is to understand the impact of equity constraint \eqref{eq:eq_8} on the shutoff decisions by varying its parameters $\beta_{\kappa\nu}$. For ease of exposition, we adopt a uniform $\beta$ for any pair of non-emergency load blocks. The wildfire safety threshold is fixed to be $\epsilon = 95\%$ and the emergency load blocks remain to be blocks 1-5. We next examine the shutoff decisions by adjusting $\beta$ values, for which the number of shutoffs at each block during the rolling horizon is illustrated in Fig.~\ref{fig:DS_3}. 
To begin with, we set $\beta = \infty$, which indicates that the ratio in \eqref{eq:eq_8} is not rigorously constrained. The results in Fig.~\ref{fig:DS_3_1} reveal that only 29 out of 52 non-emergency load blocks participate in the load shedding operations, with varying numbers of shutoffs among these blocks. Afterward, we set $\beta = 6$, and thus the ratio ${\sum (1-z_{\kappa,t})}/{\sum (1-z_{\nu,t})}, \: t\in \cal T$ for any two blocks is contained within the range of $\left[\frac{1}{6}, 6\right]$.
Compared with the benchmark case ($\beta = \infty$), this modification ensures the active participation of all non-emergency load blocks in contributing to the shutoff plans over the rolling horizon, as illustrated in Fig.~\ref{fig:DS_3_2}. As we continue to reduce $\beta$ values to 4 and 2, the range of ${\sum (1-z_{\kappa,t})}/{\sum (1-z_{\nu,t})}$ becomes even more constrained. The variance of load shedding occurrences is reduced as $\beta$ gets smaller, leading to a more even-handed shutoff strategy, as shown in Fig.~\ref{fig:DS_3_3} and Fig.~\ref{fig:DS_3_4}. Meanwhile, with a more balanced allocation of the total load shedding task among load blocks, the maximum number of shutoffs for any block is also reduced from 6 to 4.

Note that pursing a completely unbiased strategy (e.g., $\beta = 1$) may not be the most favorable choice, as this often results in uneconomical decisions and can encounter infeasibility occasionally. The main point we aim to convey with this test case is that by fine-tuning the parameters in constraint \eqref{eq:eq_8}, the equity among the associated load blocks can be conveniently adjusted. Of course, going beyond the load block level, the constraint can be further extended to involve the comparison of shutoffs between any two areas.


\section{Conclusion} \label{sec:con}
This paper presents a dynamic topology reconfiguration approach for equitable and resilient networked microgrid operations. We put forth a rolling-horizon optimization formulation to determine the optimal microgrid network topology during each time interval, in an effort to mitigate the wildfire risk. To foster a more equitable load shutoff plan, we additionally impose equity constraints on shutoff actions over the span of the rolling horizon. The effectiveness of the proposed algorithm has been demonstrated in mitigating biased and unjust load shedding decisions among different groups of communities.
This research also reveals new opportunities and paves the way for further studies in the area. These may include explicit design of equity constraints for groups with different backgrounds such as age, income, and preparedness (e.g., \cite{wang2023local,dugan2023social}), and microgrid planning and operation problems in the presence of intermittent and uncertain renewables.

\bibliography{bibliography}





\bibliographystyle{IEEEtran}

\itemsep2pt


\end{document}